\documentclass[prl,showpacs,twocolumn]{revtex4}
\usepackage{graphicx}
\usepackage{dcolumn}
\usepackage{bm}
\begin{document}
\title{ Curvature-induced anisotropic spin-orbit splitting in carbon nanotubes}

\author{ L. Chico,
M. P. L\'opez-Sancho, and M.C. Mu\~noz} 
\affiliation{
Instituto de Ciencia de Materiales de Madrid, Consejo Superior de
Investigaciones Cient{\'{\i}}ficas, Cantoblanco, 28049 Madrid, Spain}

\date{\today}

\begin{abstract}

We have theoretically explored the spin-orbit interaction in carbon nanotubes. 
We show that, besides the dependence on chirality
and diameter, the effects of spin-orbit coupling are anisotropic: 
spin splitting is larger for the higher valence or the lower electron band depending on the specific 
tube. Different tube behaviors can be grouped in three families, according to the so-called chiral index. 
Curvature-induced changes in the orbital hybridization have a crucial role, and they are shown to be family-dependent.
Our results explain recent experimental results which have evidenced the importance 
of spin-orbit effects in carbon nanotubes.

\noindent
\end{abstract}

\maketitle

Improvements in the quality of carbon nanotubes (CNTs) have enabled 
the fabrication
of quantum dots aiming at the realization of spintronics devices 
\cite{SK05,LFLY06,HP07,MM08}. 
CNTs present a high Fermi velocity and a twofold 
orbital degeneracy originating from the topology of the honeycomb lattice.
The unique fourfold degeneracy of CNTs energy states (spin plus orbital moment)
has been observed in CNT quantum dots (QDs) by magnetic field spectroscopy measurements \cite {JH05}
and makes them particularly interesting since, besides the spin degree of freedom, they
present the orbital moment to allow for quantum manipulation.
In a recent experiment \cite{KIRM08}, spin-orbit coupling has been directly observed in
CNT  as a splitting of the fourfold degeneracy of a single-electron energy level in ultra-clean  QDs.
This important finding seems to be in contradiction with the interpretation of earlier 
experiments in defect-free CNTs, from which independent spin and orbital symmetries and electron-hole symmetry 
have been deduced \cite{JH04}.
Besides showing the importance of spin-orbit effects in carbon nanotubes, Kuemmeth {\it et al.}  \cite{KIRM08} 
point out an unexplained anisotropic splitting of electron and holes 
in carbon nanotube quantum dots, which deserves further exploration.

On theoretical grounds, spin-orbit interaction (SOI) has been investigated on CNTs by
deriving an effective mass Hamiltonian including a weak SOI in carbon orbitals
to the lowest order in perturbation theory \cite{ando}.
Band splitting was 
 found considering surface curvature effects \cite{entin}, 
as well as in the electron spin resonance spectra 
of achiral CNTs derived by low-energy theory \cite{dimar}.
In an earlier work, we showed that
the inclusion of the full lattice symmetry is essential
for deriving  spin-orbit (SO) effects in CNTs \cite{CLM04}. Employing an empirical tight-binding model, 
we demonstrated  an intrinsic symmetry dependence of SOI effects.
As confirmed by recent experimental results \cite{KIRM08},
we showed that, in the absence of a magnetic field, 
CNTs present spin-orbit split bands at the Fermi level.
In addition, SOI induces zero-field spin splitting in chiral CNTs, while
Kramers theorem on time-reversal symmetry alongside the inversion symmetry
preserve the spin-degeneracy in
 achiral---{\it i.e.}, $(n,n)$ armchair and $(n,0)$ zigzag---nanotubes \cite{notation}.
More recent works \cite{HHGB06} have indicated the importance of curvature in the
SOI effects investigated with a continuum model by perturbation theory, reporting 
the appearance of a gap and spin-splitting in the CNT band spectrum due to SOI.
Here we show that, besides the dependence on the diameter and
chirality, SOI effects in CNTs exhibit an electron-hole anisotropy
which is specific to the tube. Different nanotube behaviors with respect to SOI are grouped into three families, 
which have also arisen with respect to other electronic properties \cite{CBR07}.

We have performed electronic structure calculations using 
an empirical tight-binding Hamiltonian
including a four orbital
$sp^3$ basis set  \cite{SK,TL}.
The CNT unit cell is formed by rolling up a portion of
a graphene sheet; thus, the actual discrete nature of the lattice and 
curvature effects are taken into account. 
The atomic spin-orbit interaction term included in the Hamiltonian
is given by
${ H_{SO}} = \lambda {\bf L}\cdot{\bf  S}$, 
where $\lambda$ is a renormalized atomic SO coupling constant and {\bf L} and {\bf S} stand for
the orbital  and  spin  angular momentum of the electron, respectively.
The spin quantization direction has been chosen parallel to the carbon nanotube axis.
Different estimations of $\lambda$ have been done \cite{ando,HHGB06}, always assuming
a very small value, considerably reduced in graphite/graphene and CNTs 
with respect to the value for atomic C ($\approx 12$ meV). 
However, the recent experiments of Ref. \cite{KIRM08} point out an enhancement of the role of SOI in CNTs
with respect to that of graphene; thus, the exact value of the SO coupling parameter is still under discussion. 
In the present work, SO-induced energy splittings are given 
relative to the strength of the SOI, and only in band structure results we have chosen an artificially large value of $\lambda$, for the sake of clarity in the Figures.

\begin{figure}
\includegraphics[width=\columnwidth,clip]{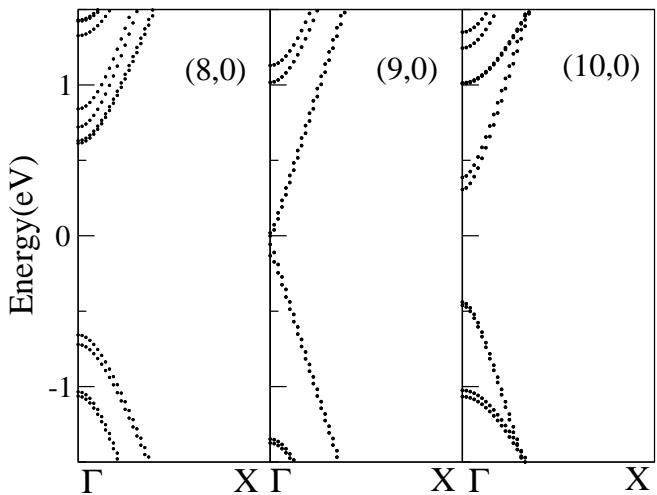}
\caption{
\label{b89y10}
Band structures calculated around the Fermi level
for three zigzag  CNTs:  (8,0) tube from the $(3q-1,0)$ family,  (9,0) primary metallic nanotube from the $(3q,0)$ family, and
 (10,0) tube from the $(3q+1,0)$ series. Band structures
have been calculated including the SOI term with $\lambda=0.2$ eV.
}
\end{figure}

We focus first on achiral zigzag nanotubes with a chiral angle $\phi = 0^{\rm o}$,
where curvature effects are expected to be largest \cite{curv,kleiner}. 
Zigzag $(n,0)$ tubes can be classified as primary metals \cite{kleiner} if $n=3q$, $q$ being an integer, or semiconducting, if $n=3q \pm 1$.
Without curvature effects, primary metallic zigzag tubes have a band crossing 
at the Fermi level; this occurs at the 
center of the Brillouin zone (BZ). The bands  crossing at  $\Gamma$
are fourfold degenerate, and inclusion of curvature opens a small gap. 
Zigzag tubes with $n=3q\pm 1$ 
are semiconducting
with the bandgap at $\Gamma$; the top valence and bottom conduction bands are also fourfold degenerate. 
Inclusion of SO interaction partially removes the band degeneracy
at $E_F$, although the split bands remain spin-degenerate. 
The energy splittings induced by the SO interaction term are different
for the highest valence band (VB) and the lowest conduction band (CB): in two of the
families, for CNTs $(3q,0)$ and $(3q-1,0)$, the splitting is larger for the
VB, whereas for NTs $(3q+1,0)$ the splitting is larger for the CB. 
In Figure \ref{b89y10} we show one particular example of each of the three 
zigzag families, namely the (8,0), (9,0) and (10,0) CNTs. Although our analysis concentrates 
on those bands closest to the Fermi level, it can be seen that the splitting of all other bands
is also anisotropic and specific to each band. 
In particular, note that the second CB and VB show
opposite behaviors to those of the bottom CB and top VB bands (see below).

\begin{figure}
\includegraphics[width=\columnwidth,clip]{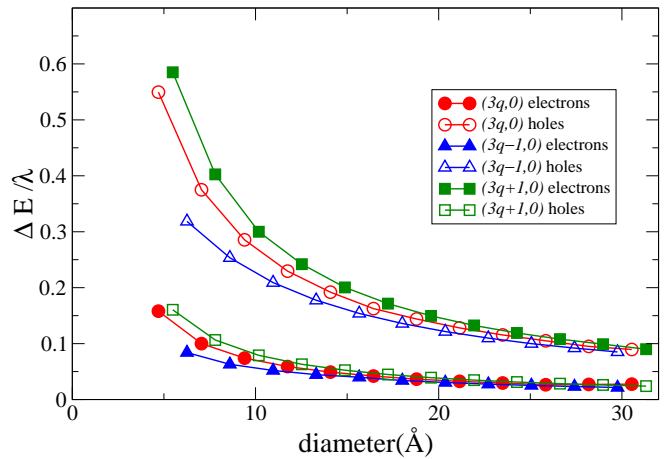}
\caption{\label{fign0}
(Color online) 
Normalized spin-orbit splitting of the top valence and bottom conduction bands
 as a function of the tube diameter for the three zigzag families.}
\end{figure}

In Figure \ref{fign0} the SO energy splitting of the top valence and bottom conduction bands 
is represented as a function
of the tube diameter for the three zigzag CNT series, showing the family behavior described before.
The curvature effect is clearly seen: the SO energy splitting 
is much higher for CNTs with the smallest diameters;
it can be noted as well how 
the differences among the 
three series decreases with increasing diameter. We have checked that the splitting 
is not negligible for larger NTs: for example, the (80,0) tube,  with a diameter of 62.64 \AA, roughly the same size 
of the NT measured in \cite{KIRM08}, has a top VB normalized splitting of 0.04.
An analogous family behavior of zigzag CNTs has been observed
in other physical properties, such as the band gaps \cite{YM95,SDD00,ando05}, and it has 
been related to the trigonal warping effect.
The three families can be understood by resorting to the zone folding approach.
Within this approximation, the energy bands of a general $(n,m)$ CNT are given by imposing 
periodic boundary conditions to the graphene sheet: if the lines of allowed $k$ vectors touch the 
Fermi points ($K$ or $K'$) of the hexagonal graphene BZ, the nanotube is a metal. 
This occurs when $n-m=3q$, with $q$ integer; otherwise it is a semiconductor. 
The two semiconductor families, $n-m=3q \pm 1$, correspond to nanotubes for which the quantization 
lines  
yielding the energy gap are at opposite sides of the Fermi point $K$ (or $ K'$). 
Thus, CNTs can be classified into 
three families: $n-m=3q+\nu$, with $\nu = 0,\pm 1$ being the so-called chiral index, hereafter 
referred as family index \cite{family}.
Denoting as $K$ the BZ special point with coordinates $(\frac{4\pi}{3a},0)$, 
the $\nu=-1$ has the closest quantization line yielding the gap to its left, over the 
$\Gamma K$ line,
 whereas the $\nu=+1$ family has it to the right, over the $KM$ line. 
Of course, if we choose the $K'$ Fermi point  to classify the nanotube families, the relative positions (right/left)  of the quantization lines yielding the gap are reversed with respect to $K$. For simplicity,  henceforth we will refer our discussion to the $K$ point with coordinates given above.

\begin{figure}
\includegraphics[width=\columnwidth,clip=]{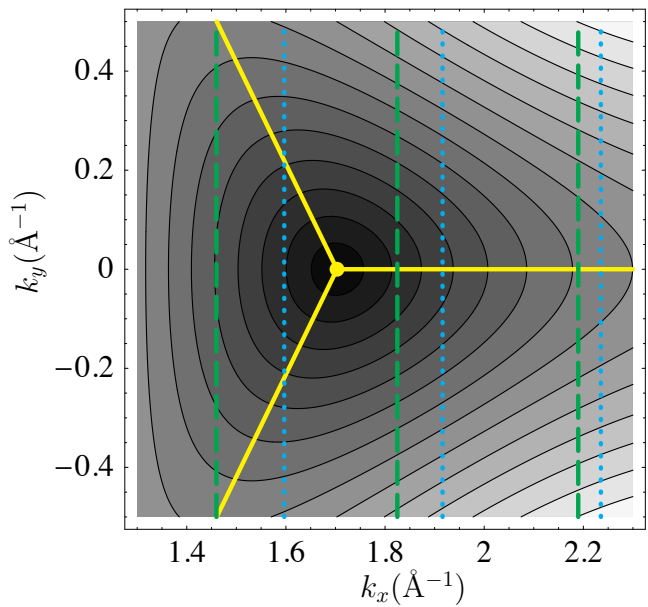}
\caption{\label{figcontyellow} (Color online) 
Contour plot of the graphene bandstructure calculated around the Fermi  $K$ point.
The quantization lines corresponding to the (7,0) (green, dashed) and (8,0) (blue, dotted) CNTs closer to this point are also shown. }
\end{figure}

In Fig. \ref{figcontyellow} we illustrate this general classification for the case of zigzag CNTs, with two examples belonging to different series. 
The allowed quantization lines of the (7,0) and
(8,0) tubes closer to $K$ are shown over the graphene $\pi$-band structure energy contour plot.
The trigonal warping effect is clearly seen in these contours:
quantization lines at opposite sides of $K$ correspond to graphene
bands with appreciably different slopes, although they are at similar distances.
As the $\pi$ graphene bands crossing at $K$ have different symmetry, 
the character of the conduction band  changes from antibonding along the $\Gamma K$ line to bonding along the $KM$ line. This explains why tubes belonging to $\nu = \pm 1$ families show an opposite behavior in their conduction (or valence) bands: the two bands closest to the Fermi energy have a different symmetry. In addition, it clarifies why in a given tube, as described above, the second VB (CB) has a reverse behavior with respect to the first VB (bottom CB): these second bands arise from quantization lines at opposite sides of $K$ with respect to the first.
Furthermore, the similar behavior of the metallic and the semiconductor $\nu = -1$ families can 
be explained by noticing that curvature effects shift the Fermi $K$ point in Fig. 
\ref{figcontyellow} to the right, in such a way that the allowed
quantization lines of the metallic tubes closer to $K$ fall onto the same side as those 
corresponding to the $\nu=-1$ family \cite {kleiner}.

On the other hand, 
curvature changes the hybridization of the orbitals. 
The linear graphene bands crossing at $K$ are of a pure $\pi$ character around $E_F$; however, 
rehybridization of $\sigma$ and $\pi$ orbitals
is very important for small-radii nanotubes, and non-negligible in general.
In order to show
the $\sigma$-$\pi$ hybridization of the zigzag nanotubes we have calculated the
contribution of each of the four orbitals forming the basis set 
for the  valence and conduction band states closest in energy to the Fermi level 
at  $\Gamma$; the results summed over all the atoms of the unit cell are shown in Table \ref{tab:hor}. 
NTs belonging to the $\nu = +1$ family 
have a larger contribution of the $\sigma$
orbitals---that is, a larger $\sigma$-$\pi$ rehybridization---in the conduction band than in the valence band; however, in 
tubes of the $\nu = 0, -1$ families the $\sigma$ orbital density is larger in the valence band. 
Spin-orbit effects are more important for bands with larger curvature-induced rehybridization:
 therefore, curvature effects are responsible for the observed electron-hole anisotropic SO splitting.
The different mixing of $\pi$ and $\sigma$ orbitals,
due to the curvature of the tubes,
was shown to affect the energy gaps of semiconducting
nanotubes \cite{YM95} with a similar family dependence. 
Here we have shown that it also influences the value of spin-orbit splitting in 
CNTs. 

\begin{table}[htbp]
   \centering
\caption{Electronic densities} %
   \begin{tabular}{c @{\extracolsep{12pt}} c c@{\extracolsep{12pt}} c c @{\extracolsep{12pt}}c c } %
     \hline\hline
       & \multicolumn{2}{c}{(8,0)} & \multicolumn{2}{c}{(9,0)} & \multicolumn{2}{c}{(10,0)} \\
  orbital  & CB & VB  & CB & VB  & CB  & VB \\ \hline
     $\sigma$  & 0.057 & 0.087 & 0.058 & 0.103 & 0.121 & 0.061 \\
     $\pi$         & 0.943 & 0.913 & 0.941 & 0.897 & 0.879 & 0.939 \\
  \hline\hline
  \end{tabular}
   \label{tab:hor}
\end{table}

We consider now chiral NTs $(n,m)$, with $n \ne m\ne 0$ that do not have an
inversion center.
As mentioned above, curvature effects induce a shift of the Fermi wave vector $k_F$,
opening a small gap at the Fermi energy
in the primary metallic chiral CNTs, $(m-n)=3q$. But, in contrast to the results shown for achiral zigzag  NTs, 
in both metal and semiconductor chiral NTs the SO interaction lifts all degeneracies.

\begin{figure}
\includegraphics[width=\columnwidth,clip]{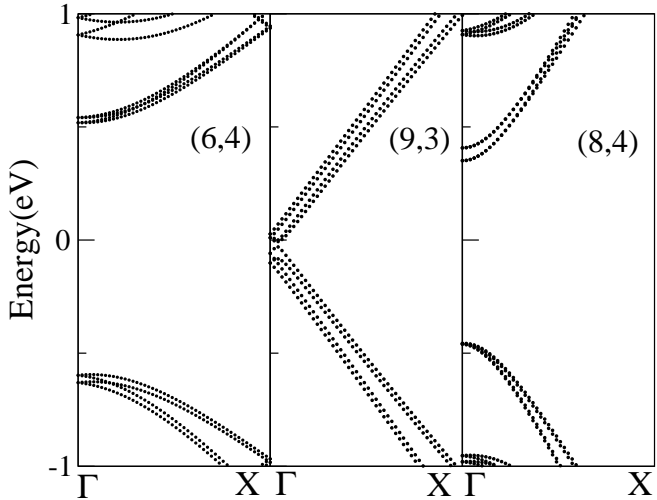}
\caption{
\label{b3chiral}
Band structures calculated with SO interaction around the Fermi level
for three chiral  CNTs: the (6,4) tube from the $\nu = -1$ family, 
the (9,3) primary metallic nanotube, and
the (8,4) tube from the $\nu = +1$ family. }
\end{figure}

\begin{figure}
\includegraphics[width=\columnwidth,clip]{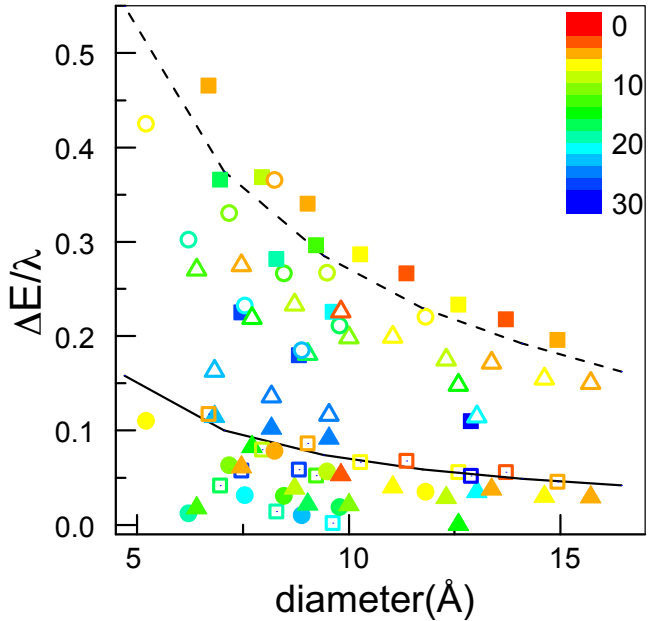}
\caption{ \label{splitchiral} (Color) 
Normalized energy band splittings for chiral CNTs versus diameter. Squares stand for $\nu=+1$ family, triangles for $\nu-1$  and circles for $\nu=0$ ; full (open) indicate the top VB (bottom CB) splittings. Symbol color indicates the nanotube chiral angle. Full (dashed) line: Bottom CB (top VB) splittings for metallic zigzag tubes.}
\end{figure}

The calculated energy splittings for these tubes follow the same behavior
as those obtained for achiral zigzag CNTs: for tubes with $\nu=0$ ($n-m=3q$) and $\nu=-1$ ($n-m=3q-1$)
the energy
splitting is larger for the highest VB, while for tubes with $\nu=+1$ ($n-m=3q+1$) 
the splitting is larger for the lowest CB. 
As an example for chiral tubes, the band structures calculated including the SOI 
for three particular tubes, (6,4), (9,3), and (8,4)
belonging to each of the three families, are shown in Fig. \ref{b3chiral} around the Fermi level.
However, chirality has an important effect in SOI:
this is illustrated in Fig. \ref{splitchiral}, where the band splittings vs. diameter for 
chiral tubes belonging to the three families are depicted. 
The symbol color indicates the NT chiral angle. As a guide to the eye, the
zigzag metallic $\nu =0$ results are also shown. It can be seen how chiral 
tubes follow the family behavior already described; notwithstanding, chirality effects
introduce deviations from the zigzag monotonic behavior,
due to the different orientation of the relevant quantization lines with respect to the $K$  point. In general, the higher the chiral angle, the larger the deviation from the behavior of zigzag nanotubes. 

In summary, we have shown that spin orbit effects in carbon nanotubes is anisotropic:
The energy splittings induced by the SO interaction term are different
for the highest valence band and the lowest conduction band depending on the tube family; 
the magnitude of the SO splitting correlates with the $\sigma$-$\pi$ hybridization induced by curvature.
These dissimilar $\sigma$-$\pi$ hybridizations of the valence and conduction band states, which in turn depend on the position 
of the quantization lines yielding the gap with respect to the graphene $K$ point,
are the reason for the experimentally observed anisotropy in SO splitting in absence of external fields.

L. C. acknowledges fruitful discussions with A. Ayuela, J. I. Cerd\'a, and A. Ruiz.
This work has been partially supported by the Spanish DGES under
grants FIS2005-05478-C02-01, FIS2008-00124, MAT2006-05122 and MAT2006-06242.



\begin{thebibliography}{99}
\bibitem{notation}
Nanotubes are labeled by the numbers $(n,m)$,
given by
the unrolled circumference vector ${\bm C}_h = n {\bm a}_1 + m {\bm a}_2 $ in the graphene lattice.
${\bm a}_1$, ${\bm a}_2$ are the two lattice vectors of graphene
at 60$^{\rm o}$; $a =|{\bm a}_i|$.
\bibitem{family} The family index $\nu$ as defined in this work, or equivalently, 
$\bmod(n-m,3)$, is called chiral index by other authors. Given that nanotubes with 
different chiralities may belong to the same family, we prefer the previous denomination.
\end{thebibliography}
\end{document}